# Mixer-First Receiver with wide-RF Range

Harshit Roy , Mrigank Sharad ,Indian Institute Of Technology,Kharagpur

*Abstract*— In the Passive Mixer first receiver, four mosfet, and four baseband impedance can synthesize High-Q bandpass filter. This on-chip High-Q bandpass filter can replace bulky, expensive and off-chip SAW filters. The impedance which is seen by the antenna is tuned by switch resistance of the Passive Mixer and the input impedance of the Trans-impedance amplifier (baseband impedance) to match antenna impedance. The gain of the feedback loop of the TIA controls the input impedance of the TIA. Further, M baseband impedance can be replaced by M/4 complex impedance. This complex impedance provides a wide RF range by changing the transconductance.

## I. INTRODUCTION

A traditional receiver consists of a bandpass filter, LNA and mixer. The in band signal is usually accompanied by out of band blocker which desensitizes a circuit(LNA) even if they do not reduce the gain the gain to zero [4]. SAW filters are used to attenuate out of band blockers. Usually, Band-select filter such as SAW filter is implemented off-chip. They degrade the receiver sensitivity and are bulky. Different Receiver architecture and circuit topologies are being proposed to replace the SAW filter. M phase passive mixer can provide on-chip High-Q bandpass filtering, and they can replace SAW filters. High programmability is the need of the next generation receiver. Since these BPF are not tunable but to provide programmability in RF band, there are multiple frontends in the receiver, and each of them is tuned to a different frequency using a matching network [1].In one of the other approaches, antenna directly interfaces to a passive mixer and impedance matching is performed by tuning the baseband impedance. Noise and Linearity performance of these type of receiver is presented[2].

In this paper, we present a similar approach in which M phase High-Q bandpass filters can be used to interface directly with an antenna. The impedance matching can be performed using Passive Mixer Transparency Property of the 4-Phase Passive Mixer. Baseband impedance in the M-phase High-Q bandpass filter can be replaced with M/4 complex filter. Passive mixer frequency translates the complex impedance to LO frequency [2]. Transconductance ($G_m$)-stage (in the complex impedance) can control the offset of the center frequency of High-Q bandpass filter from LO frequency [2]. The control over the center frequency of the bandpass filter provides programmability in selecting the band of operation.

## II. PASSIVE MIXER TRANSPARENCY

4-phase passive mixer consists of four switches, which are successively turned on by 25% duty cycle non-overlapping periodic clocks. Each switch samples the RF voltage onto four capacitors ($C_L$) loaded by the baseband resistors $R_B$. Since LO-clocks are completely non-overlapping i.e. one switch is on at a time, switch resistance $R_{SW}$ can be lumped with antenna impedance $R_a$.

Effective Antenna Impedance $R_a' = R_a + R_{SW}$

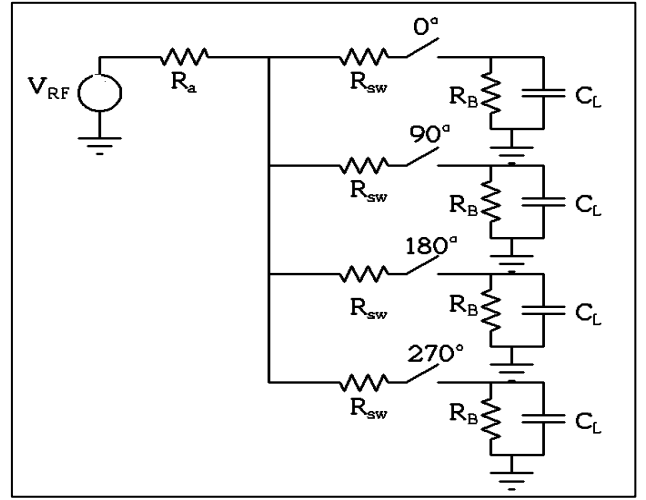

**Figure 1:** 4-phase Passive Mixer with baseband Impedance ($R_B || C_L$)

The impedance (at $\omega \approx \omega_{LO}$) looking into the 4-phase passive mixture is modeled as switch resistance in series with the parallel combination of $R_{sh}$ and $\gamma R_b$. The complete analysis of the input impedance is presented in one of the reference paper [2]. The input impedance is tuned by changing the $R_{SW}$ and baseband-impedance $R_b$. If $R_b$ is very small, $R_{in}$ approaches $R_{sw}$. If the $R_b$ is large compared to shunt impedance $R_{sh}$, the effect of baseband impedance cannot be seen in $R_{in}$. Impedance matching ability of the 4 phase passive mixer with $S_{11}$ analysis has been presented in one of the paper [2].

$$\gamma = \frac{2}{\pi^2} \approx 0.203$$

$$R_{sh} = R_a' \frac{4\gamma}{1-4\gamma} \approx 4.3 R_a'$$

Further in this paper, $R_B$ will be replaced with the complex impedance ($Z_B$). We have to keep the $R_{sh}$ comparable or higher than $\gamma Z_B$ to observe the effect of the change in $Z_B$ on the input impedance.

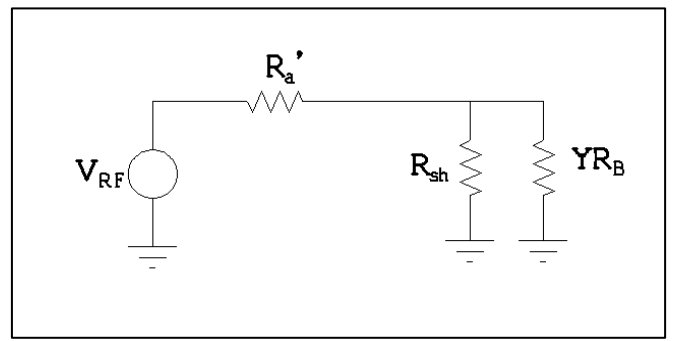

**Figure 2:** LTI Equivalent circuit for the Passive Mixer

## III. COMPLEX IMPEDANCE

A complex impedance $Z_C(\omega)$ is defined as an impedance that receives a complex current $I_{Re}(t) + jI_{Im}(t)$ and produces a complex response voltage equal to $[I_{Re}(t) + jI_{Im}(t)] \times Z_C(t)$. Fourier transform of complex impedance $Z_C(\omega)$ is asymmetrical around DC. In this paper, we replace baseband impedance by a low-Q bandpass complex impedance whose center frequency is shifted by $\omega_{OC}$ from the DC.

A complex impedance can be realized using baseband impedance and a voltage-dependent current source [3]. The output voltage is dependent on the input current by the following relation.

$$V_{Re}(\omega) + jV_{Im}(\omega) = \frac{Z_{BB}(\omega)}{1 + jG_m(\omega)Z_{BB}(\omega)}[I_{Re}(\omega) + jI_{Im}(\omega)]$$

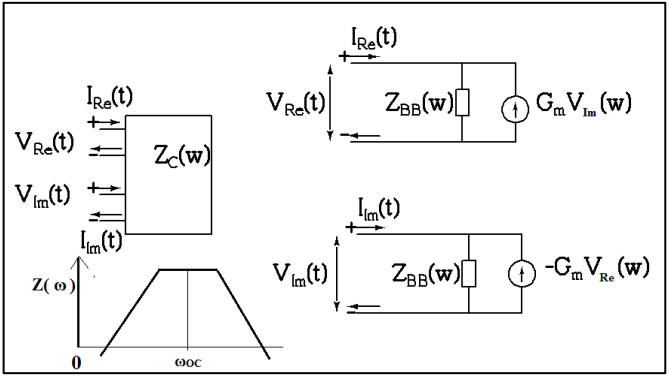

**Figure 3: Complex Impedance**

If the baseband impedance is made of a resistor $R_{BB}$ and a capacitance $C_{BB}$ in parallel combination, the above equation modifies to the below equation. The equation means that the low. pass transfer function is shifted in frequency domain to the left by $G_m/C_{BB}$. As long as the bandwidth of transconductor is much larger than $G_m/C_{BB}$ (shift in the frequency response), the limited bandwidth of the $G_m$ stage is not a concern [3].

$$Z_C(\omega) = \frac{R_{BB}}{1 + jR_{BB}C_{BB}\left(\omega + \frac{G_m}{C_{BB}}\right)}$$

*A. Simulation Result*

We did PSS ($\omega_{LO}$ = 1GHz and $\omega_{RF}$ = 1.01GHz) analysis followed by PAC analysis for the Passive Mixer with Complex Impedance with variable transconductance in 0.18μm Technology. In this case, we varied the transconductance by varying the current. Later, we will propose another way to vary the transconductance of the transconductance stage.

At very high and low frequency from the local oscillator frequency, the Input impedance is equal to switch resistance because $Z_{BB}(\omega)$ will be approximately equal to 0 at those frequency. In this simulation, we have used switches of size 100μ/0.18μ to keep the switch resistance minimum ($\omega_{LO}≈5\Omega$). Switch charge Injection will be significant due to the large W, and it will be input dependent. In the fully differential design of the M-phase filter, the constant portion of the channel charge would appear as common-mode, posing no threat [3].

On increasing the current, the transconductance of the NMOS increases. Center frequency of the Input Impedance shifts by $G_m/(2\pi C_{BB})$ from the local oscillator frequency as expected. Baseband capacitance used in the simulation result is 55pF.

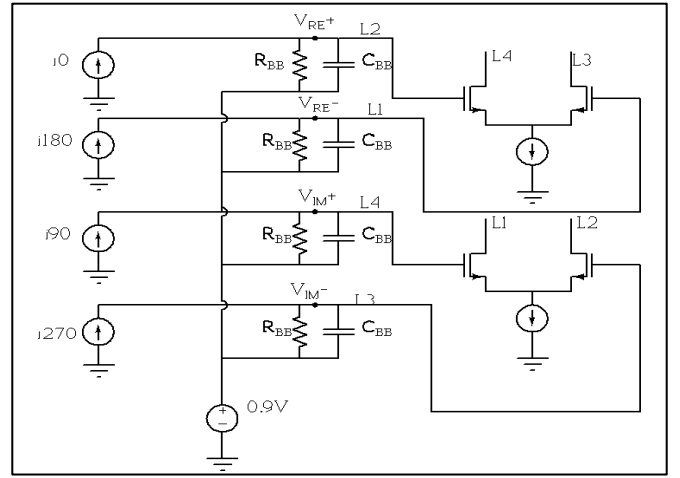

**Figure 4: Realization of the Complex Impedance with $Z_{BB}$ = $R_B \parallel C_B$ and transconductance stage using NMOS**

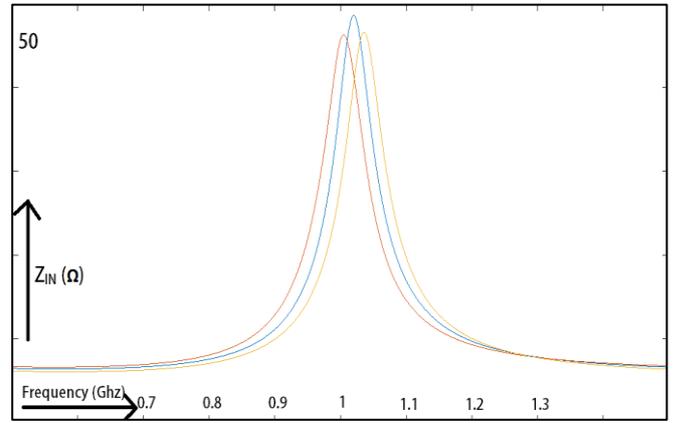

**Figure 5: Input Impedance of the Passive Mixer ($Z_{IN}$) with varying transconductance ($G_m$)**

IV. TRANSIMPEDANCE AMPLIFIER (TIA)

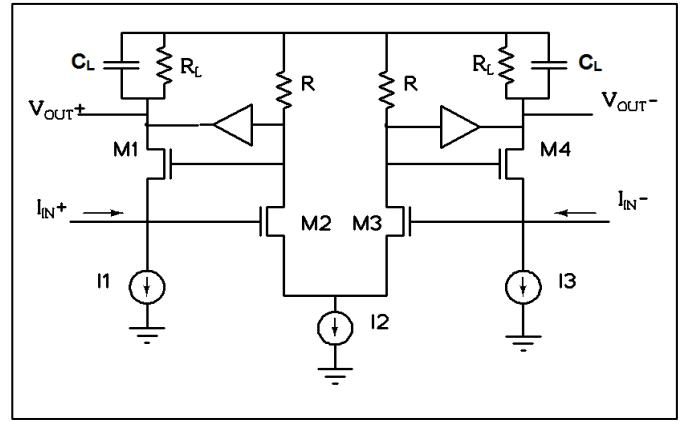

**Figure 6: Regulated Cascode Transimpedance Amplifier**

Regulated Cascode (RGC) Topology of Transimpedance Amplifier consists of a common-gate amplifier and common-source amplifier as active feedback. Input Impedance of the common gate amplifier is decreased due to negative feedback.

Input Impedance of the RGC is given by

$$R_{IN} = \frac{\left(R_L \parallel \frac{1}{j\omega C_L}\right) + r_{01}}{1 + r_{01}g_{m1}(g_{m2}(r_{02} \parallel R) + 1)}$$

The zero in the input impedance can be pushed towards a higher frequency by using a buffer in the feedback loop. The input of the RGC is connected with baseband capacitance ($C_{BB}$ = 55pF) to give low pass frequency response as shown in Figure 7. Power Consumption of RGC is 0.18mW.

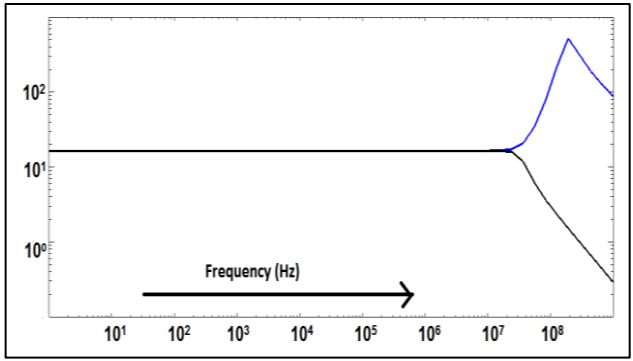

**Figure 7: Input Impedance of RGC without and with baseband Capacitance $C_{BB}$**

In the final architecture (Figure 10) , there is 4-phase passive mixer followed by a complex impedance, which consists of RGC amplifier and an NMOS as transconductance stage. The current of the transconductance stage has to be varied by a significant amount to get an appreciable change in Gm. Therefore, we will amplify the signal by Differential Input Differential Output Amplifier follow by NMOS in a differential configuration to realize the transconductance stage. Amplifier has to be of sufficient bandwidth. Instead of varying the current of the NMOS, we can now vary the gain of the amplifier stage or current of the NMOS to vary the transconductance.

*A. Simulation Results*

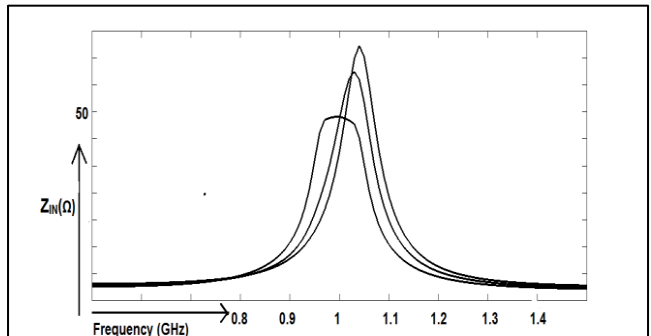

**Figure 8: Input Impedance of the Passive Mixer (TIA as baseband resistance) with the varying transconductance**

Input Impedance of Passive Mixer should match the Antenna Impedance at the required frequency. To get Input Impedance as a function of frequency, we did PSS Analysis with the Beat frequency of 10 MHz followed by PAC Analysis (0.5 - 1.5 GHz with input Amplitude of 1mV). We observed a shift in the center frequency of the bandpass filter. There is an increase in the Input Impedance which can be adjusted to 50ohm by changing the switch resistance. A switch can be implemented in multiple sets, and some of them can be turned on/off to adjust the Input Impedance to 50ohm.

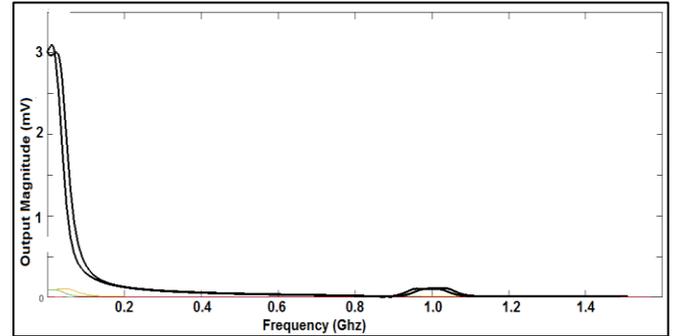

**Figure 9: Output of TIA**

Figure 9 contains the output voltage of TIA (real) for the -1 and 1 sideband of Local Oscillator (LO). We observe a gain of 3. For sinusoidal input (1mV and 1.01 GHz), the negative frequency will be shifted toward DC by +1 sideband of LO and the positive frequency will be moved toward DC by -1 sideband of LO.

$$f(out) = f(in) + K_i \times PSS_{fund}$$

$K_i$ are the sidebands and $f_{in}$ represents the input frequency.

*B. Noise*

We did PSS analysis followed by PNOISE analysis. Output Spectral Noise Density is 25.76nV/Hz. Noise Performance of the Passive Mixer has been described in one of the reference paper [2].

## V. CONCLUSION

In this paper, we have introduced Regulated Cascode (RGC) as Transimpedance Amplifier (TIA). RGC can provide low input impedance with low power consumption. We have provided a transistor level realization of complex Impedance with the output noise. The idea of complex Impedance and passive mixer transparency can be combined to give mixer-first receiver, which has wide RF-tuning.

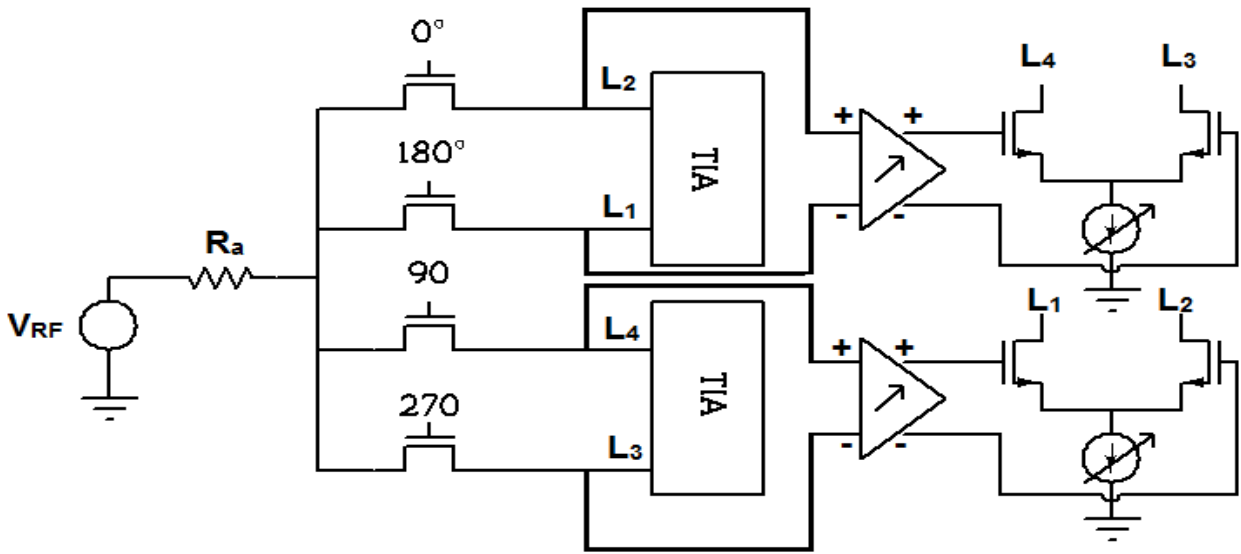

**Figure 10 : Complete Architecture**